\documentstyle[prl,multicol,aps,graphics]{revtex}
\begin{document}
\draft
\title{Optimization and Parallelization of a force field for silicon using OpenMP}
\author{Stefan Goedecker}
\address{D\'epartement de recherche fondamentale sur la mati\`ere condens\'ee,\\
         SP2M/NM, CEA-Grenoble, 38054 Grenoble cedex~9, France}
\date{\today}
\maketitle
\begin{abstract}
The force field by Lenosky and coworkers is the latest force field
for silicon which is one of the most studied materials. It has turned
out to be highly accurate in a large range of test cases. The
optimization and parallelization
of this force field using OpenMp and Fortan90 is described here.
The optimized program allows us to handle a very large number
of silicon atoms in large scale simulations. Since all the parallelization
is hidden in a single subroutine that returns the total energies and forces,
this subroutine can be called from within a serial program in an user friendly 
way.
\end{abstract}

\section{Program survey}

Title of program: siliconiap

Computer hardware and operating system: Any shared memory 
computer running under Unix or Linux

Programming Language: Fortran90 with OpenMP compiler directives

Memory requirements: roughly 150 words per atom

No. of bits in a word: 64

No. of processors used: tested on up to 4 processors

Has the code been vectorized or parallelized: Parallelized withe OpenMP

No. of bytes in distributed program, including test data, etc: 50 000

Distribution format: Compressed tar file

Keywords: Silicon, Interatomic potential, Force field, Molecular dynamics

Nature of physical problem: condensed matter physics

Method of solution: Interatomic potential

Restrictions on the complexity of the problem: None

Typical running time: 30 $\mu$sec per step and per atom on a Compaq DEC Alpha

Unusual features of the program: None

\section{Introduction}
Due to its technological importance, silicon is one of the most studied 
materials. For small system sizes ab-initio density functional
 calculations~\cite{kohn}  
are the preferred approach. Unfortunately this kind of calculation 
becomes unfeasible for larger systems required to study problems such as 
interfaces or extended defects. For this type of calculations one resorts 
to force fields which are several orders of magnitude faster. Recent 
progress in the development of force fields has demonstrated that they 
can be a reliable tool for such studies. A highly accurate silicon 
force field has been developed by Lenosky and coworkers~\cite{lenos}. 
Its transferability has been demonstrated by extensive tests containing 
both bulk and cluster systems~\cite{lenos}. Its accuracy is in part due 
to the fact that second nearest 
neighbor interactions are included. This makes it unfortunately 
somewhat slower than force fields containing only nearest neighbor 
interactions. In the following a highly optimized parallel implementation 
of this force field will be presented that allows large scale calculations 
with this force field. The parallelization is achieved by using OpenMP, 
an emerging industry standard for medium size shared memory 
parallel computers. 

Molecular dynamics calculations~\cite{md1} have also been parallelized 
on distributed memory supercomputers~\cite{lomdahl}. This approach is 
considerably more complex than the one presented here. Since few 
researches have access to massively parallel supercomputers and are 
willing to overcome the complexities of doing molecular dynamics 
on such machines, medium scale parallelization~\cite{french} of 
molecular dynamics has an important place in practice.

\section{Calling the subroutine}
User friendliness was one of the major design goals in the development 
of this routine. Using Fortran90 made it possible to hide all the 
complexities in an object oriented fashion from the user. The calling 
sequence is just

\begin{verbatim}
call lenosky(nat,alat,rxyz,fxyz,ener,coord,ener_var,coord_var,count)
\end{verbatim}
On input the user has to specify the number of atoms, $nat$, the 
vector $alat$ containing the 3 lattice constant of the orthorhombic periodic volume 
and the atomic positions $rxyz$. The program then returns the total energy, 
$ener$, the forces, $fxyz$, the average coordination number, the variation of the 
energy per atom and of the coordination number as well as an counter that 
is increased in each call. In particular the user has not to supply any 
Verlet list.

Since the calculation of the forces is typically much more expensive than 
the update of the atomic positions in molecular dynamics or geometry 
optimizations, we expect that the subroutine will be called in most cases 
from within a serial program. In case the user is on a shared memory machine 
the subroutine will then nevertheless be executed in parallel if the 
program is compiled with the appropriate OpenMP options.

In addition the subroutine can of course also be used on a serial machine.
In this case all the parallelization directives are considered by the 
compiler to be comments.

\section{Calculation of the Verlet list}
The Verlet list gives all the atoms that are contained within 
the potential cutoff distance $cut$ of any given atom.
Typically the Verlet list consists of two integer arrays.
The first array, called $lsta$ in this work, points to the first/last 
neighbor position in the second array $lstb$ that contains the numbering 
of the atoms that are neighbors.
A straightforward implementation for a non-periodic system containing $nat$ atoms 
is shown below. In this simple case the search through all atoms 
is sequential with respect to their numbering and it is redundant to give both the 
starting positions $lsta(1,iat)$ and the ending position $lsta(2,iat)$, 
since $lsta(1,iat)=lsta(2,iat-1)+1$.  
But in the more complicated linear scaling algorithm to be presented below, 
both will be needed. 

\begin{verbatim}
        indc=0
      do 10 iat=1,nat
c       starting position
        lsta(1,iat)=indc+1
        do 20 jat=1,nat
        if (jat.ne.iat) then
          xrel1= rxyz(1,jat)-rxyz(1,iat)
          xrel2= rxyz(2,jat)-rxyz(2,iat)
          xrel3= rxyz(3,jat)-rxyz(3,iat)
          rr2=xrel1**2 + xrel2**2 + xrel3**2
           if ( rr2 .le. cut**2 ) then
           indc=indc+1
c          nearest neighbor numbers
           lstb(indc)=jat
           endif
        endif
20      continue
c       ending position
        lsta(2,iat)=indc
10      continue

\end{verbatim}

This  straightforward implementations has a quadratic scaling 
with respect to the numbers of atoms. Due to this scaling the calculation 
of the Verlet list starts to dominate the linear scaling 
calculation of the 
energies and forces for system sizes of more than 10 000 atoms. 
It is therefore good practice to calculate the Verlet list with 
a modified algorithm that has linear scaling~\cite{md1,md2} as well.

To do this one first subdivides the system into boxes that have a side length 
that is equal to or larger than $cut$ and then finds all the atoms 
that are contained in each box. The CPU time for this first step is 
less than 1 percent of the entire Verlet list calculation. 
Hence this part was not parallelized. It could significantly affect the parallel 
performance according to Amdahls law~\cite{mybook} only if more 
than 50 processors are used. The largest SMP machines at our disposal 
had however only 4 processors.

\begin{figure}[ht]
     \begin{center}
      \setlength{\unitlength}{1cm}
       \begin{picture}( 5.,7.)           % figure dimensions
        \put(-1.5,0.5){\includegraphics{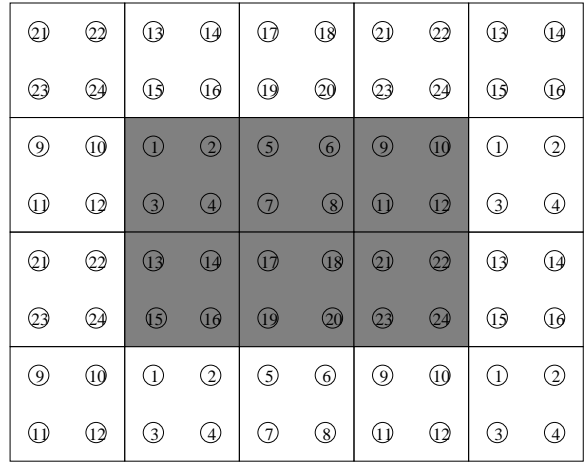}}   % VAX
       \end{picture}
 \caption[]{\label{period}
           Illustration of the construction of the cell structure 
           necessary for a linear scaling calculation of the nearest 
           neighbor list for a 2-dimensional case. The periodic volume 
           is indicated by the dark background. The bright cells are 
           replicated dark cells. }
      \end{center}
     \end{figure}

To implement periodic boundary conditions all atoms within a distance 
$cut$ of the boundary of the periodic volume are replicated on the opposite 
part as shown in Figure~\ref{period}. This part is 
equally well less than 1 percent of the CPU time for the Verlet 
list for a 8000 atom system. Being a surface term it becomes even 
smaller for larger systems. Consequently it wasn't parallelized either.

After these two preparing steps one has to search only among all the atoms 
in the reference cell containing the atom for which one wants to find 
its neighbors as well as all the atoms in the cells 
neighboring this reference cell (26 cells in 3
dimensions). This implies that starting and ending positions $lsta$ for 
the atoms $1$ to $nat$ are not calculated in a sequential way, necessitating,  
as mentioned before, separate starting and ending positions in the 
array $lsta$. The corresponding parallel 
code is shown below. The indices $l1,l2,l3$ refer to the cells, 
$icell(l1,l2,l3,0)$ contains the number of atoms in cell $l1,l2,l3$ 
and $icell(l1,l2,l3,*)$ their numbering. The array $rel$ saves the relative 
positions and distances that will again be needed in the loop calculating 
the forces and energies. Each thread has its own starting position 
$iam*myspace+1$ in the shared memory space $lstb$ and these starting 
positions are uniformly distributed. This approach allows the different 
threads to work independently. The resulting speedup is much higher 
than the one that one would obtain by calculating in the parallel version 
an array $lstb$ that is identical to the one from the serial version. 
If there are on the average more neighbors than expected (24 by default) 
the allocated space becomes too small. In this case the 
array $lstb$ is deallocated and a new larger version is allocated.
This check for sufficient memory requires some minimal amount of 
coordination among the processors and is implemented by a critical section. 
If a reallocation is necessary, a message is written into an file to alert 
the user of the inefficiency 
due to the need of a second calculation of the Verlet list. 
\begin{verbatim}
        allocate(lsta(2,nat))
        nnbrx=24
2345    nnbrx=3*nnbrx/2
        allocate(lstb(nnbrx*nat),rel(5,nnbrx*nat))

        indlstx=0

!$omp parallel  &
!$omp private(iat,cut2,iam,ii,indlst,l1,l2,l3,myspace,npr) &
!$omp shared (indlstx,nat,nn,nnbrx,ncx,ll1,ll2,ll3,icell,lsta,lstb,lay, & 
!$omp rel,rxyz,cut,myspaceout)


        npr=1
!$       npr=omp_get_num_threads()
        iam=0
!$       iam=omp_get_thread_num()

        cut2=cut**2
        myspace=(nat*nnbrx)/npr  
        if (iam.eq.0) myspaceout=myspace
! Verlet list, relative positions
        indlst=0
      do 6000,l3=0,ll3-1
      do 6000,l2=0,ll2-1
      do 6000,l1=0,ll1-1
      do 6600,ii=1,icell(0,l1,l2,l3)
        iat=icell(ii,l1,l2,l3)
        if ( ((iat-1)*npr)/nat .eq. iam) then
        lsta(1,iat)=iam*myspace+indlst+1
        call sublstiat(iat,nn,ncx,ll1,ll2,ll3,l1,l2,l3,myspace, & 
             rxyz,icell,lstb(iam*myspace+1),lay,rel(1,iam*myspace+1),cut2,indlst)
        lsta(2,iat)=iam*myspace+indlst
        endif

6600    continue
6000    continue
!$omp critical
        indlstx=max(indlstx,indlst)
!$omp end critical
!$omp end parallel  

           if (indlstx.gt.myspaceout) then
               write(10,*) count,'NNBRX too  small', nnbrx
               deallocate(lstb,rel)
               goto 2345
           endif



        subroutine sublstiat(iat,nn,ncx,ll1,ll2,ll3,l1,l2,l3,myspace, & 
                   rxyz,icell,lstb,lay,rel,cut2,indlst)
        implicit real*8 (a-h,o-z)
        dimension rxyz(3,nn),lay(nn),icell(0:ncx,-1:ll1,-1:ll2,-1:ll3), &
                  lstb(0:myspace-1),rel(5,0:myspace-1)

        do 6363,k3=l3-1,l3+1
        do 6363,k2=l2-1,l2+1
        do 6363,k1=l1-1,l1+1
        do 6363,jj=1,icell(0,k1,k2,k3)
          jat=icell(jj,k1,k2,k3)
          if (jat.eq.iat) goto 6363
          xrel= rxyz(1,iat)-rxyz(1,jat)
          yrel= rxyz(2,iat)-rxyz(2,jat)
          zrel= rxyz(3,iat)-rxyz(3,jat)  
          rr2=xrel**2 + yrel**2 + zrel**2
          if ( rr2 .le. cut2 ) then
           indlst=min(indlst,myspace-1)
           lstb(indlst)=lay(jat)
           tt=sqrt(rr2)
           tti=1.d0/tt
           rel(1,indlst)=xrel*tti
           rel(2,indlst)=yrel*tti
           rel(3,indlst)=zrel*tti
           rel(4,indlst)=tt
           rel(5,indlst)=tti
           indlst= indlst+1
          endif
6363        continue

        return
        end
\end{verbatim}

\section{Calculation of the energies and forces}
The computationally most important part taking some 80 percent of the CPU time 
is the calculation of the energies and forces. The energy expression for 
the Lenosky force field is given by
\begin{equation}
E = \sum_{i,j} \phi(r_{ij}) + \sum_i U \left( \sum_j \rho(r_{ij}) + 
    \sum_{j,k} f(r_{ij}) f(r_{ik}) g(cos(\theta_{jik})) \right) \label{energy}
\end{equation}
All the functions ($\phi$, $U$, $\rho$, $f$, $g$) in this energy expression
are given by cubic splines. The subroutine for evaluating the cubic spline 
is listed below. The case that the argument is outside the cubic spline interval 
$[tmin,tmax]$ is rare and unimportant for performance considerations.
The important cubic spline case is characterized by many dependencies.
In the case of such dependencies the latency of the functional unit 
pipeline comes into play and reduces the attainable speed~\cite{mybook}. 
A latency of some 20 cycles comes from the first two statements 
( tt=(x-tmin)*hi ; klo=tt) alone, requiring arithmetic operations and 
a floating point to integer conversion. For this reason the calculation of tt 
was taken out of the (most likely occurring) else block to overlap 
its evaluation with the evaluation of the if clauses.
To further speed up the evaluation of the splines the structure of the 
energy expression~\ref{energy} was exploited. In the computationally 
most important loop over $j$ and $k$ two splines ($f(r_{ik})$ and 
$g(cos(\theta{jik}))$ ) have to be evaluated. Inlining by hand the 
subroutine splint for both evaluations and calculating alternatingly 
one step of the first spline evaluation and one step of the second 
spline evaluation introduces two independent streams. This reduces the 
effect of latencies and boosts speed. Compilers are not able to 
do these complex type of optimizations. The best 
performance after these optimizations was obtained with low level 
compiler optimization flags (-O3 -qarch=pwr3 -qtune-pwr3 on IBM Power3, 
-O2 on the Compaq DEC Alpha, -O2 -xW on Intel Pentium4 )

\begin{verbatim}
        subroutine splint(ya,y2a,tmin,tmax,hsixth,h2sixth,hi,n,x,y,yp)
        implicit real*8 (a-h,o-z)
        dimension y2a(0:n-1),ya(0:n-1)

! interpolate if the argument is outside the cubic spline interval [tmin,tmax]
        tt=(x-tmin)*hi
        if (x.lt.tmin) then
          yp=hi*(ya(1)-ya(0)) -  &
          ( y2a(1)+2.d0*y2a(0) )*hsixth
          y=ya(0) + (x-tmin)*yp
        else if (x.gt.tmax) then
          yp=hi*(ya(n-1)-ya(n-2)) +  &
          ( 2.d0*y2a(n-1)+y2a(n-2) )*hsixth
          y=ya(n-1) + (x-tmax)*yp
! otherwise evaluate cubic spline
        else
          klo=tt
          khi=klo+1
          ya_klo=ya(klo)
          y2a_klo=y2a(klo)
          b=tt-klo
          a=1.d0-b
          ya_khi=ya(khi)
          y2a_khi=y2a(khi)
          b2=b*b
          y=a*ya_klo
          yp=ya_khi-ya_klo
          a2=a*a
          cof1=a2-1.d0
          cof2=b2-1.d0
          y=y+b*ya_khi
          yp=hi*yp
          cof3=3.d0*b2
          cof4=3.d0*a2
          cof1=a*cof1
          cof2=b*cof2
          cof3=cof3-1.d0
          cof4=cof4-1.d0
          yt1=cof1*y2a_klo
          yt2=cof2*y2a_khi
          ypt1=cof3*y2a_khi
          ypt2=cof4*y2a_klo
          y=y + (yt1+yt2)*h2sixth
          yp=yp + ( ypt1 - ypt2 )*hsixth
        endif
      return
      end

\end{verbatim}

The final single processor performance for the entire subroutine is 
460 Mflops on a Compaq DEC Alpha at 833 MHz, 300 Mflops on a IBM Power3 
at 350 MHz and 550 Mflops on a Pentium 4. 
In order to obtain a high parallel speedup in this 
central part of the subroutine the threads are completely decoupled.
This was done by introducing private copies for each thread to 
accumulate the energies $tener$ and forces $txyz$. The global energy 
and force are summed up in an additional loop at the end of the parallel 
region in a critical section. 

\begin{verbatim}
!$omp parallel  &
!$omp private(iam,npr,iat,iat1,iat2,lot,istop,tcoord,tcoord2, & 
!$omp tener,tener2,txyz,f2ij,f3ij,f3ik,npjx,npjkx) &
!$omp shared (nat,nnbrx,lsta,lstb,rel,ener,ener2,fxyz,coord,coord2,istopg)

        npr=1
!$       npr=omp_get_num_threads()
        iam=0
!$       iam=omp_get_thread_num()

        npjx=300 ; npjkx=3000
        istopg=0

        if (npr.ne.1) then 
! PARALLEL CASE
! create temporary private scalars for reduction sum on energies and 
!        temporary private array for reduction sum on forces
!$omp critical
        allocate(txyz(3,nat),f2ij(3,npjx),f3ij(3,npjkx),f3ik(3,npjkx))
!$omp end critical
       if (iam.eq.0) then
        ener=0.d0
        ener2=0.d0
        coord=0.d0
        coord2=0.d0
        do 121,iat=1,nat
        fxyz(1,iat)=0.d0
        fxyz(2,iat)=0.d0
121     fxyz(3,iat)=0.d0
       endif

        lot=nat/npr+.999999999999d0
        iat1=iam*lot+1
        iat2=min((iam+1)*lot,nat)
        call subfeniat(iat1,iat2,nat,lsta,lstb,rel,tener,tener2, &
               tcoord,tcoord2,nnbrx,txyz,f2ij,npjx,f3ij,npjkx,f3ik,istop)
!$omp critical
        ener=ener+tener
        ener2=ener2+tener2
        coord=coord+tcoord
        coord2=coord2+tcoord2
        istopg=istopg+istop
        do 8093,iat=1,nat
        fxyz(1,iat)=fxyz(1,iat)+txyz(1,iat)
        fxyz(2,iat)=fxyz(2,iat)+txyz(2,iat)
        fxyz(3,iat)=fxyz(3,iat)+txyz(3,iat)
8093    continue
!$omp end critical
        deallocate(txyz,f2ij,f3ij,f3ik)

        else
! SERIAL CASE
        iat1=1
        iat2=nat
        allocate(f2ij(3,npjx),f3ij(3,npjkx),f3ik(3,npjkx))
        call subfeniat(iat1,iat2,nat,lsta,lstb,rel,ener,ener2, &
               coord,coord2,nnbrx,fxyz,f2ij,npjx,f3ij,npjkx,f3ik,istop)
        deallocate(f2ij,f3ij,f3ik)

        endif
!$omp end parallel

        if (istopg.gt.0) stop 'DIMENSION ERROR (see WARNING above)'
        ener_var=ener2/nat-(ener/nat)**2 
        coord=coord/nat
        coord_var=coord2/nat-coord**2 

        deallocate(rxyz,icell,lay,lsta,lstb,rel)

        end


        subroutine subfeniat(iat1,iat2,nat,lsta,lstb,rel,tener,tener2, &
               tcoord,tcoord2,nnbrx,txyz,f2ij,npjx,f3ij,npjkx,f3ik,istop)
        implicit real*8 (a-h,o-z)
        dimension lsta(2,nat),lstb(nnbrx*nat),rel(5,nnbrx*nat),txyz(3,nat)
        dimension f2ij(3,npjx),f3ij(3,npjkx),f3ik(3,npjkx)

     initialize data ........

! create temporary private scalars for reduction sum on energies and
        tener=0.d0
        tener2=0.d0
        tcoord=0.d0
        tcoord2=0.d0
        istop=0
        do 121,iat=1,nat
        txyz(1,iat)=0.d0
        txyz(2,iat)=0.d0
121     txyz(3,iat)=0.d0

! calculation of forces, energy

        do 1000,iat=iat1,iat2

        dens2=0.d0
        dens3=0.d0
        jcnt=0
        jkcnt=0
        coord_iat=0.d0
        ener_iat=0.d0
        do 2000,jbr=lsta(1,iat),lsta(2,iat)
        jat=lstb(jbr)
        jcnt=jcnt+1
        if (jcnt.gt.npjx) then 
            write(6,*) 'WARNING: enlarge npjx'  
            istop=1
        endif

        fxij=rel(1,jbr)
        fyij=rel(2,jbr)
        fzij=rel(3,jbr)
        rij=rel(4,jbr)
        sij=rel(5,jbr)

! coordination number calculated with soft cutoff between first and  
! second nearest neighbor
        if (rij.le.2.36d0) then
        coord_iat=coord_iat+1.d0
        else if (rij.ge.3.83d0) then
        else
        x=(rij-2.36d0)*(1.d0/(3.83d0-2.36d0))
        coord_iat=coord_iat+(2*x+1.d0)*(x-1.d0)**2
        endif

! pairpotential term        
        call splint(cof_phi,dof_phi,tmin_phi,tmax_phi, &
           hsixth_phi,h2sixth_phi,hi_phi,10,rij,e_phi,ep_phi)
        ener_iat=ener_iat+(e_phi*.5d0)
        txyz(1,iat)=txyz(1,iat)-fxij*(ep_phi*.5d0)
        txyz(2,iat)=txyz(2,iat)-fyij*(ep_phi*.5d0)
        txyz(3,iat)=txyz(3,iat)-fzij*(ep_phi*.5d0)
        txyz(1,jat)=txyz(1,jat)+fxij*(ep_phi*.5d0)
        txyz(2,jat)=txyz(2,jat)+fyij*(ep_phi*.5d0)
        txyz(3,jat)=txyz(3,jat)+fzij*(ep_phi*.5d0)

! 2 body embedding term        
        call splint(cof_rho,dof_rho,tmin_rho,tmax_rho, &
             hsixth_rho,h2sixth_rho,hi_rho,11,rij,rho,rhop)
        dens2=dens2+rho
        f2ij(1,jcnt)=fxij*rhop
        f2ij(2,jcnt)=fyij*rhop
        f2ij(3,jcnt)=fzij*rhop

! 3 body embedding term        
        call splint(cof_fff,dof_fff,tmin_fff,tmax_fff, & 
             hsixth_fff,h2sixth_fff,hi_fff,10,rij,fij,fijp)

        do 3000,kbr=lsta(1,iat),lsta(2,iat)
        kat=lstb(kbr)
        if (kat.lt.jat) then
        jkcnt=jkcnt+1
        if (jkcnt.gt.npjkx) then 
            write(6,*) 'WARNING: enlarge npjkx' 
            istop=1
        endif

! begin optimized version
        rik=rel(4,kbr)
      if (rik.gt.tmax_fff) then
        fikp=0.d0 ; fik=0.d0
        gjik=0.d0 ;  gjikp=0.d0 ; sik=0.d0
        costheta=0.d0 ; fxik=0.d0 ; fyik=0.d0 ; fzik=0.d0
      else if (rik.lt.tmin_fff) then
        fxik=rel(1,kbr)
        fyik=rel(2,kbr)
        fzik=rel(3,kbr)
        costheta=fxij*fxik+fyij*fyik+fzij*fzik
        sik=rel(5,kbr)
        fikp=hi_fff*(cof_fff(1)-cof_fff(0)) -  & 
             ( dof_fff(1)+2.d0*dof_fff(0) )*hsixth_fff
        fik=cof_fff(0) + (rik-tmin_fff)*fikp
        tt_ggg=(costheta-tmin_ggg)*hi_ggg
        if (costheta.gt.tmax_ggg) then
           gjikp=hi_ggg*(cof_ggg(8-1)-cof_ggg(8-2)) + & 
                ( 2.d0*dof_ggg(8-1)+dof_ggg(8-2) )*hsixth_ggg
                 gjik=cof_ggg(8-1) + (costheta-tmax_ggg)*gjikp
        else
           klo_ggg=tt_ggg
           khi_ggg=klo_ggg+1
           cof_ggg_klo=cof_ggg(klo_ggg)
           dof_ggg_klo=dof_ggg(klo_ggg)
           b_ggg=tt_ggg-klo_ggg
           a_ggg=1.d0-b_ggg
           cof_ggg_khi=cof_ggg(khi_ggg)
           dof_ggg_khi=dof_ggg(khi_ggg)
           b2_ggg=b_ggg*b_ggg
           gjik=a_ggg*cof_ggg_klo
           gjikp=cof_ggg_khi-cof_ggg_klo 
           a2_ggg=a_ggg*a_ggg
           cof1_ggg=a2_ggg-1.d0
           cof2_ggg=b2_ggg-1.d0
           gjik=gjik+b_ggg*cof_ggg_khi
           gjikp=hi_ggg*gjikp
           cof3_ggg=3.d0*b2_ggg
           cof4_ggg=3.d0*a2_ggg
           cof1_ggg=a_ggg*cof1_ggg
           cof2_ggg=b_ggg*cof2_ggg
           cof3_ggg=cof3_ggg-1.d0
           cof4_ggg=cof4_ggg-1.d0
           yt1_ggg=cof1_ggg*dof_ggg_klo
           yt2_ggg=cof2_ggg*dof_ggg_khi
           ypt1_ggg=cof3_ggg*dof_ggg_khi
           ypt2_ggg=cof4_ggg*dof_ggg_klo
           gjik=gjik + (yt1_ggg+yt2_ggg)*h2sixth_ggg
           gjikp=gjikp + ( ypt1_ggg - ypt2_ggg )*hsixth_ggg
        endif
      else
        fxik=rel(1,kbr)
        tt_fff=rik-tmin_fff
        costheta=fxij*fxik
        fyik=rel(2,kbr)
        tt_fff=tt_fff*hi_fff
        costheta=costheta+fyij*fyik
        fzik=rel(3,kbr)
        klo_fff=tt_fff
        costheta=costheta+fzij*fzik
        sik=rel(5,kbr)
        tt_ggg=(costheta-tmin_ggg)*hi_ggg
        if (costheta.gt.tmax_ggg) then
          gjikp=hi_ggg*(cof_ggg(8-1)-cof_ggg(8-2)) + & 
                ( 2.d0*dof_ggg(8-1)+dof_ggg(8-2) )*hsixth_ggg
                gjik=cof_ggg(8-1) + (costheta-tmax_ggg)*gjikp
          khi_fff=klo_fff+1
          cof_fff_klo=cof_fff(klo_fff)
          dof_fff_klo=dof_fff(klo_fff)
          b_fff=tt_fff-klo_fff
          a_fff=1.d0-b_fff
          cof_fff_khi=cof_fff(khi_fff)
          dof_fff_khi=dof_fff(khi_fff)
          b2_fff=b_fff*b_fff
          fik=a_fff*cof_fff_klo
          fikp=cof_fff_khi-cof_fff_klo 
          a2_fff=a_fff*a_fff
          cof1_fff=a2_fff-1.d0
          cof2_fff=b2_fff-1.d0
          fik=fik+b_fff*cof_fff_khi
          fikp=hi_fff*fikp
          cof3_fff=3.d0*b2_fff
          cof4_fff=3.d0*a2_fff
          cof1_fff=a_fff*cof1_fff
          cof2_fff=b_fff*cof2_fff
          cof3_fff=cof3_fff-1.d0
          cof4_fff=cof4_fff-1.d0
          yt1_fff=cof1_fff*dof_fff_klo
          yt2_fff=cof2_fff*dof_fff_khi
          ypt1_fff=cof3_fff*dof_fff_khi
          ypt2_fff=cof4_fff*dof_fff_klo
          fik=fik + (yt1_fff+yt2_fff)*h2sixth_fff
          fikp=fikp + ( ypt1_fff - ypt2_fff )*hsixth_fff
         else
              klo_ggg=tt_ggg
              khi_ggg=klo_ggg+1
           khi_fff=klo_fff+1
              cof_ggg_klo=cof_ggg(klo_ggg)
           cof_fff_klo=cof_fff(klo_fff)
              dof_ggg_klo=dof_ggg(klo_ggg)
           dof_fff_klo=dof_fff(klo_fff)
              b_ggg=tt_ggg-klo_ggg
           b_fff=tt_fff-klo_fff
              a_ggg=1.d0-b_ggg
           a_fff=1.d0-b_fff
              cof_ggg_khi=cof_ggg(khi_ggg)
           cof_fff_khi=cof_fff(khi_fff)
              dof_ggg_khi=dof_ggg(khi_ggg)
           dof_fff_khi=dof_fff(khi_fff)
              b2_ggg=b_ggg*b_ggg
           b2_fff=b_fff*b_fff
              gjik=a_ggg*cof_ggg_klo
           fik=a_fff*cof_fff_klo
              gjikp=cof_ggg_khi-cof_ggg_klo 
           fikp=cof_fff_khi-cof_fff_klo 
              a2_ggg=a_ggg*a_ggg
           a2_fff=a_fff*a_fff
              cof1_ggg=a2_ggg-1.d0
           cof1_fff=a2_fff-1.d0
              cof2_ggg=b2_ggg-1.d0
           cof2_fff=b2_fff-1.d0
              gjik=gjik+b_ggg*cof_ggg_khi
           fik=fik+b_fff*cof_fff_khi
              gjikp=hi_ggg*gjikp
           fikp=hi_fff*fikp
              cof3_ggg=3.d0*b2_ggg
           cof3_fff=3.d0*b2_fff
              cof4_ggg=3.d0*a2_ggg
           cof4_fff=3.d0*a2_fff
              cof1_ggg=a_ggg*cof1_ggg
           cof1_fff=a_fff*cof1_fff
              cof2_ggg=b_ggg*cof2_ggg
           cof2_fff=b_fff*cof2_fff
              cof3_ggg=cof3_ggg-1.d0
           cof3_fff=cof3_fff-1.d0
              cof4_ggg=cof4_ggg-1.d0
           cof4_fff=cof4_fff-1.d0
              yt1_ggg=cof1_ggg*dof_ggg_klo
           yt1_fff=cof1_fff*dof_fff_klo
              yt2_ggg=cof2_ggg*dof_ggg_khi
           yt2_fff=cof2_fff*dof_fff_khi
              ypt1_ggg=cof3_ggg*dof_ggg_khi
           ypt1_fff=cof3_fff*dof_fff_khi
              ypt2_ggg=cof4_ggg*dof_ggg_klo
           ypt2_fff=cof4_fff*dof_fff_klo
              gjik=gjik + (yt1_ggg+yt2_ggg)*h2sixth_ggg
           fik=fik + (yt1_fff+yt2_fff)*h2sixth_fff
              gjikp=gjikp + ( ypt1_ggg - ypt2_ggg )*hsixth_ggg
           fikp=fikp + ( ypt1_fff - ypt2_fff )*hsixth_fff
         endif
      endif
! end optimized version

        tt=fij*fik
        dens3=dens3+tt*gjik

        t1=fijp*fik*gjik
        t2=sij*(tt*gjikp)
        f3ij(1,jkcnt)=fxij*t1 + (fxik-fxij*costheta)*t2
        f3ij(2,jkcnt)=fyij*t1 + (fyik-fyij*costheta)*t2
        f3ij(3,jkcnt)=fzij*t1 + (fzik-fzij*costheta)*t2

        t3=fikp*fij*gjik
        t4=sik*(tt*gjikp)
        f3ik(1,jkcnt)=fxik*t3 + (fxij-fxik*costheta)*t4
        f3ik(2,jkcnt)=fyik*t3 + (fyij-fyik*costheta)*t4
        f3ik(3,jkcnt)=fzik*t3 + (fzij-fzik*costheta)*t4
        endif
3000        continue
2000        continue
        dens=dens2+dens3
        call splint(cof_uuu,dof_uuu,tmin_uuu,tmax_uuu, & 
             hsixth_uuu,h2sixth_uuu,hi_uuu,8,dens,e_uuu,ep_uuu)
        ener_iat=ener_iat+e_uuu

! Only now ep_uu is known and the forces can be calculated, lets loop again
        jcnt=0
        jkcnt=0
        do 2200,jbr=lsta(1,iat),lsta(2,iat)
        jat=lstb(jbr)
        jcnt=jcnt+1
        txyz(1,iat)=txyz(1,iat)-ep_uuu*f2ij(1,jcnt)
        txyz(2,iat)=txyz(2,iat)-ep_uuu*f2ij(2,jcnt)
        txyz(3,iat)=txyz(3,iat)-ep_uuu*f2ij(3,jcnt)
        txyz(1,jat)=txyz(1,jat)+ep_uuu*f2ij(1,jcnt)
        txyz(2,jat)=txyz(2,jat)+ep_uuu*f2ij(2,jcnt)
        txyz(3,jat)=txyz(3,jat)+ep_uuu*f2ij(3,jcnt)
        
! 3 body embedding term        
        do 3300,kbr=lsta(1,iat),lsta(2,iat)
        kat=lstb(kbr)
        if (kat.lt.jat) then
        jkcnt=jkcnt+1

        txyz(1,iat)=txyz(1,iat)-ep_uuu*(f3ij(1,jkcnt)+f3ik(1,jkcnt))
        txyz(2,iat)=txyz(2,iat)-ep_uuu*(f3ij(2,jkcnt)+f3ik(2,jkcnt))
        txyz(3,iat)=txyz(3,iat)-ep_uuu*(f3ij(3,jkcnt)+f3ik(3,jkcnt))
        txyz(1,jat)=txyz(1,jat)+ep_uuu*f3ij(1,jkcnt)
        txyz(2,jat)=txyz(2,jat)+ep_uuu*f3ij(2,jkcnt)
        txyz(3,jat)=txyz(3,jat)+ep_uuu*f3ij(3,jkcnt)
        txyz(1,kat)=txyz(1,kat)+ep_uuu*f3ik(1,jkcnt)
        txyz(2,kat)=txyz(2,kat)+ep_uuu*f3ik(2,jkcnt)
        txyz(3,kat)=txyz(3,kat)+ep_uuu*f3ik(3,jkcnt)


        endif
3300        continue
2200        continue

        tener=tener+ener_iat
        tener2=tener2+ener_iat**2
        tcoord=tcoord+coord_iat
        tcoord2=tcoord2+coord_iat**2

1000    continue

        return
        end

\end{verbatim}

In addition to the energy and the forces the program still returns 
the coordination number as well as the variance of the energy 
per atom and the coordination number. The coordination number 
is calculated using a soft cutoff between the first and second nearest 
neighbor distance. These extra calculations are very cheap and not visible 
as an increase in the CPU time

\section{Parallel performance results}
Table~\ref{speedup_dec} shows the final overall speedups obtained by the program. 
The results were obtained for an 8000 atom system, but the 
CPU time per call and atom is nearly independent of system size.

\begin{table}[ht]
\caption[]{ Timings in $\mu$sec for a combined evaluation of 
the forces and the energy per 
particle as well as the corresponding speedups (in parentheses) 
on an IBM SP3 based on a 375 MHz Power3 processor, 
on a Compaq SC 232  based on a 833Mhz EV67 processor and  
on an Intel Pentium4 biprocessor at 2 GHz 
 \label{speedup_dec}}
\begin{tabular}{|l||c|c|c|} \hline
 number of processors & IBM Power3 & Compaq EV67 & Intel P4 \\ \hline 
  1 & 46   &  30       &  25         \\ \hline
  2 & 24 (1.9)  & 16 (1.9)  & 13 (1.9)    \\ \hline
  4 & 13 (3.5)  & 8.6 (3.5) &             \\ \hline
  8 & 7.7 (6.0) &           &             \\ 
\end{tabular}
\end{table}

Obtaining such high speedups was not straightforward. Only the 
Compaq Fortran90 compiler was able to use in the original version 
of the program the OpenMP 'parallel do' directive to obtain a good speedup.
Both the IBM compiler and the Intel compiler failed. In order to 
get the performances of Table~\ref{speedup_dec}, it was necessary to 
encapsulate the workload of the different threads into the subroutines 
'sublstias' and 'subfen', which amounts to doing the parallelization 
quasi by hand. Using allocatable arrays in connection with OpenMP
turned also out to be tricky. Because of these problems, the 
parallelization was much more painful that one might expect for 
a shared memory model.

\section{Conclusions}
The results show that simulations for very large silicon systems are 
feasible on relatively cheap serial or parallel computers accessible to a large 
number of researches.

I thank Tom Lenosky for his help in the implementation of the interatomic 
potential and Cyrus Umrigar for providing me with the timings on the IBM 
machine.

\end{document}